\documentclass[12pt]{article}

\usepackage{amssymb}

\usepackage[dvips]{graphicx}

\begin{document}

\title
{Towards proof of new identity for Green functions in $N=1$
supersymmetric electrodynamics.}

\author{K.V.Stepanyantz\thanks{E-mail:$stepan@theor.phys.msu.su$}\\
{\small{\em Moscow State University, physical faculty,
department of theoretical physics,}}\\
{\small{\em $119992$, Moscow, Russia}}}

\maketitle

\begin{abstract}
For the $N=1$ supersymmetric massless electrodynamics, regularized
by higher derivatives, we describe a method, by which one can try
to prove the new identity for the Green functions, which was
proposed earlier. Using this method we show that some contribution
to the new identity are really 0.
\end{abstract}


\section{Introduction.}
\hspace{\parindent}

Investigation of quantum corrections in supersymmetric theories is
an interesting and sometimes nontrivial problem. For example, in
theories with the $N=1$ supersymmetry it is possible to suggest
\cite{NSVZ_Instanton} a form of the $\beta$-function exactly to
all orders of the perturbation theory. For the $N=1$
supersymmetric electrodynamics, which is considered in this paper,
this $\beta$-function (that is called the exact Novikov, Shifman,
Vainshtein and Zakharov (NSVZ) $\beta$-function) is

\begin{equation}\label{NSVZ_Beta}
\beta(\alpha) = \frac{\alpha^2}{\pi} \Big(1-\gamma(\alpha)\Big),
\end{equation}

\noindent where $\gamma(\alpha)$ is the anomalous dimension of the
matter superfield. Obtaining such a $\beta$-function by methods of
the perturbation theory appears unexpectedly complicated, although
numerous verifications by explicit calculations up to the
four-loop approximation \cite{ThreeLoop123} confirm this
hypothesis. In these papers the $\beta$-function, defined as the
derivative of divergence in the $\overline{MS}$-scheme, was
calculated with the dimensional reduction. The result is that if
the subtraction scheme is tuned by a special way, then it is
possible to obtain the exact NSVZ $\beta$-function. Nevertheless,
it is unclear, in what scheme such a $\beta$-function is obtained.
The answer to this question in the three-loop approximation was
given in Ref. \cite{3LoopHEP}. According to this paper, the NSVZ
$\beta$-function coincides with the Gell-Mann--Low function. We
note that the most convenient method for the calculations is the
higher covariant derivative regularization \cite{Slavnov}. Unlike
the dimensional regularization \cite{tHooft}, it does not break
supersymmetry, and, unlike the dimensional reduction
\cite{Siegel}, it is not inconsistent. Using the higher derivative
regularization allows revealing an interesting feature of the
quantum correction structure in supersymmetric theories, which was
first noted in Ref. \cite{3LoopHEP}: All integrals defining the
Gell-Mann--Low function are integrals of total derivatives and can
be easily calculated. A similar feature takes place in non-Abelian
supersymmetric theories \cite{3ident,2LoopYM}. (The calculations
were made with a version of the higher derivative regularization,
breaking the BRST-invariance and supplemented by a special
subtraction scheme, which guarantees fulfilling the
Slavnov--Taylor identities \cite{Slavnov1234}.) This feature was
partially explained in Refs. \cite{SD,SDYM}. According to these
papers, substituting solutions of the Slavnov--Taylor identities
into the Schwinger--Dyson equations, it is possible to obtain the
exact $\beta$-function if we suppose existence of a new identity
for Green functions. The explicit calculations up to the four-loop
approximation \cite{Pimenov,3ident} confirm this identity. A way
of proving the new identity was proposed in Ref. \cite{Identity}.
However, it was based on the analysis of Feynman rules instead of
strict functional methods. Moreover, it works only for a
restricted class of diagrams. Nevertheless, an idea proposed in
\cite{Identity} can be strictly realized. This is demonstrated in
this paper. We will see that it is possible to give a functional
formulation for most equations, presented in Ref. \cite{Identity}.
The purpose of this paper is proposing of the method, which can be
used for a strict proving (or disproving) the new identity in the
massless $N=1$ supersymmetric electrodynamics. As we will see
later, using this method, it is possible to show that some
contributions to the new identity are 0.

This paper is organized as follows.

In Sec. \ref{Section_SUSY} we collect basic information about the
$N=1$ supersymmetric electrodynamics. In Sec.
\ref{Section_New_Identity} we remind, how it is possible to
calculate its $\beta$-function exactly to all orders, and also
present a functional formulation of the new identity for Green
functions, writing it as equality to 0 of some composite operators
correlator. Calculation of this correlator is described in Sec.
\ref{Section_Proof}. The results are briefly discussed in the
Conclusion. Some technical details of the calculations are
presented in the Appendix.


\section{$N=1$ supersymmetric electrodynamics.}
\hspace{\parindent}\label{Section_SUSY}

In this paper we consider the massless $N=1$ supersymmetric
electrodynamics, which is described in the superspace by the
action

\begin{equation}\label{SQED_Action}
S = \frac{1}{4e^2} \mbox{Re}\,\int d^4x\,d^2\theta\,W_a C^{ab} W_b
+ \frac{1}{4}\int d^4x\, d^4\theta\, \Big(\phi^* e^{2V}\phi
+\tilde\phi^* e^{-2V}\tilde\phi\Big).
\end{equation}

\noindent Here $\phi$ and $\tilde\phi$ are chiral matter
superfields and $V$ is a real scalar superfield, which contains
the gauge field $A_\mu$ as a component. The superfield $W_a$ is a
supersymmetric analogue of the gauge field stress tensor. In the
Abelian case it is defined by

\begin{equation}
W_a = \frac{1}{4} \bar D^2 D_a V,
\end{equation}

\noindent where $D_a$ and $\bar D_a$ are the right and left
supersymmetric covariant derivatives respectively. (In this paper
all left spinors we denote by a bar and $\bar D^2 \equiv \bar D^a
\bar D_a$. Indexes are raised and lowered by the charge conjugated
matrix.) Action (\ref{SQED_Action}) is invariant under the gauge
transformations

\begin{equation}\label{Gauge_Transformations}
V \to V - \frac{1}{2}(\lambda+\lambda^*); \qquad \phi\to
e^{\lambda}\phi;\qquad \tilde\phi\to e^{-\lambda} \tilde\phi,
\end{equation}

\noindent where $\lambda$ is an arbitrary chiral superfield.

In order to regularize model (\ref{SQED_Action}), it is possible
to add the term with the higher derivatives

\begin{equation}\label{Regularizing_Action}
S_{\Lambda} = \frac{1}{4 e^2} \mbox{Re}\int d^4x\,d^2\theta\,W_a
C^{ab} \frac{\partial^{2n}}{\Lambda^{2n}} W_b
\end{equation}

\noindent to its action. It is important that in the Abelian case,
the superfield $W^a$ is gauge invariant; hence there are the usual
(instead of covariant) derivatives in the regularizing term.

Model (\ref{SQED_Action}) can be standardly quantized. For this,
it is convenient to use the supergraph technique (described in
detail in book \cite{West}) and to fix a gauge by adding the terms

\begin{equation}\label{Gauge_Fixing}
S_{gf} = - \frac{1}{64 e^2}\int d^4x\,d^4\theta\, \Bigg(V D^2 \bar
D^2 \Big(1 + \frac{\partial^{2n}}{\Lambda^{2n}}\Big) V + V \bar
D^2 D^2 \Big(1+ \frac{\partial^{2n}}{\Lambda^{2n}}\Big) V\Bigg).
\end{equation}

\noindent After such terms are added, a part of the action
quadratic in the superfield $V$ takes the simplest form

\begin{equation}
S_{gauge} + S_{gf} = \frac{1}{4 e^2}\int d^4x\,d^4\theta\,
V\partial^2 \Big(1+ \frac{\partial^{2n}}{\Lambda^{2n}}\Big) V.
\end{equation}

\noindent In the Abelian case considered here, the diagrams
containing ghost loops are absent. It is well known that adding
the higher derivative term does not remove divergences in the
one-loop diagrams. To regularize them, it is necessary to insert
Pauli--Villars determinants in the generating functional
\cite{Slavnov_Book}.

The renormalized action for the considered model is

\begin{eqnarray}\label{Renormalized_Action}
&& S_{ren} = \frac{1}{4 e^2} Z_3(e,\Lambda/\mu)\, \mbox{Re}\int
d^4x\,d^2\theta\,W_a C^{ab} \Big(1+
\frac{\partial^{2n}}{\Lambda^{2n}}\Big) W_b
+\nonumber\\
&&\qquad\qquad\qquad\qquad\qquad +
Z(e,\Lambda/\mu)\,\frac{1}{4}\int d^4x\, d^4\theta\, \Big(\phi^*
e^{2V}\phi +\tilde\phi^* e^{-2V}\tilde\phi\Big).\qquad
\end{eqnarray}

\noindent Hence, the generating functional can be written as

\begin{equation}\label{Modified_Z}
Z = \int DV\,D\phi\,D\tilde \phi\, \prod\limits_i \Big(\det
PV(V,M_i)\Big)^{c_i}
\exp\Big(i(S_{ren}+S_{gf}+S_S+S_{\phi_0})\Big),
\end{equation}

\noindent where the renormalized action $S_{ren}$ is given by Eq.
(\ref{Renormalized_Action}) and the action for gauge fixing terms
is given by Eq. (\ref{Gauge_Fixing}). (It is convenient to replace
$e$ with the bare coupling constant $e_0$ in it that we will
suppose below.) The Pauli--Villars determinants are defined by

\begin{equation}\label{PV_Determinants}
\Big(\det PV(V,M)\Big)^{-1} = \int D\Phi\,D\tilde \Phi\,
\exp\Big(i S_{PV}\Big),
\end{equation}

\noindent where

\begin{eqnarray}
&& S_{PV}\equiv Z(e,\Lambda/\mu) \frac{1}{4} \int
d^4x\,d^4\theta\, \Big(\Phi^* e^{2V}\Phi
+\qquad\nonumber\\
&&\qquad\qquad\qquad + \tilde\Phi^* e^{-2V}\tilde\Phi \Big) +
\frac{1}{2}\int d^4x\,d^2\theta\, M \tilde\Phi \Phi +
\frac{1}{2}\int d^4x\,d^2\bar\theta\, M \tilde\Phi^* \Phi^*,\qquad
\end{eqnarray}

\noindent and the coefficients $c_i$ satisfy the conditions

\begin{equation}
\sum\limits_i c_i = 1;\qquad \sum\limits_i c_i M_i^2 = 0.
\end{equation}

\noindent Below, we assume that $M_i = a_i\Lambda$, where $a_i$
are some constants. Inserting the Pauli-Villars determinants
allows cancelling the remaining divergences in all one-loop
diagrams, including diagrams containing counterterm
insertions.

The terms with sources are written as

\begin{eqnarray}\label{Sources}
&& S_S = \int d^4x\,d^4\theta\,J V + \int d^4x\,d^2\theta\,
\Big(j\,\phi + \tilde j\,\tilde\phi \Big) + \int
d^4x\,d^2\bar\theta\, \Big(j^*\phi^* + \tilde j^*
\tilde\phi^*\Big).
\end{eqnarray}

\noindent Moreover, in generating functional (\ref{Modified_Z}),
we introduce the expression

\begin{equation}
S_{\phi_0} = \frac{1}{4}\int d^4x\,d^4\theta\,\Big(\phi_0^*\,
e^{2V} \phi + \phi^*\, e^{2V} \phi_0 + \tilde\phi_0^*\,
e^{-2V}\tilde\phi + \tilde\phi^*\, e^{-2V}\tilde\phi_0 \Big),
\end{equation}

\noindent where $\phi_0$, $\phi_0^*$, $\tilde\phi_0$, and
$\tilde\phi_0^*$ are scalar superfields that are not chiral or
antichiral. Generally, it is not necessary to introduce the term
$S_{\phi_0}$ in the generating functional, but the presence of the
parameters $\phi_0$ e.t.c. will be needed later.

In our notation, the generating functional for the connected Green
functions is written as

\begin{equation}\label{W}
W = - i\ln Z,
\end{equation}

\noindent and the effective action is obtained via the Legendre
transformation

\begin{equation}\label{Gamma}
\Gamma = W - \int d^4x\,d^4\theta\,J {\bf V} - \int
d^4x\,d^2\theta\, \Big(j\,\phi + \tilde j\,\tilde\phi \Big) - \int
d^4x\,d^2\bar\theta\, \Big(j^*\phi^* + \tilde j^* \tilde\phi^*
\Big),
\end{equation}

\noindent where the sources $J$, $j$, and $\tilde j$ should be
expressed in terms of fields ${\bf V}$, $\phi$, and $\tilde\phi$
using the equations

\begin{equation}
{\bf V} = \frac{\delta W}{\delta J};\qquad \phi = \frac{\delta
W}{\delta j};\qquad \tilde\phi = \frac{\delta W}{\delta\tilde j}.
\end{equation}

\noindent (The argument of the effective action, corresponding to
the field $V$, we denoted by the bold letter for convenience of
the notation.)

In this paper we will mostly calculate the Gell-Mann--Low
function, which is determined by dependence of the two-point Green
function on the momentum. In order to construct it, we write terms
in the effective action, corresponding to the renormalized
two-point Green function, in the form

\begin{equation}\label{D_Definition}
\Gamma^{(2)}_V = - \frac{1}{16\pi} \int
\frac{d^4p}{(2\pi)^4}\,d^4\theta\,{\bf V}(-p)\,\partial^2\Pi_{1/2}
{\bf V}(p)\, d^{-1}(\alpha,\mu/p),
\end{equation}

\noindent where $\Pi_{1/2}$ is a supersymmetric projection
operator, and $\alpha$ is a renormalized coupling constant. The
Gell-Mann--Low function, denoted by $\beta(\alpha)$, is defined by

\begin{equation}\label{Gell-Mann-Low_Definition}
\beta\Big(d(\alpha,\mu/p)\Big) = \frac{\partial}{\partial \ln p}
d(\alpha,\mu/p).
\end{equation}

\noindent It is well known that the Gell-Mann--Low function is
scheme independent.

The anomalous dimension is defined similarly. First we consider
the two-point Green function for the matter superfield

\begin{equation}\label{Renormalized_Gamma_2}
\Gamma^{(2)}_\phi = \frac{1}{4} \int
\frac{d^4p}{(2\pi)^4}\,d^4\theta\,\Big(\phi^*(-p,\theta)\,\phi(p,\theta)
+ \tilde\phi^*(-p,\theta)\,\tilde\phi(p,\theta) \Big) \,
G_{ren}(\alpha,\mu/p),
\end{equation}

\noindent where

\begin{equation}
G_{ren}(\alpha,\mu/p) = Z(\alpha_0,\Lambda/\mu)
G(\alpha_0,\Lambda/p),
\end{equation}

\noindent and $Z$ denotes the renormalization constant for the
matter superfield. Then the anomalous dimensions is defined by

\begin{equation}
\gamma\Big(d(\alpha,\mu/p)\Big) = -\frac{\partial}{\partial\ln p}
\ln G_{ren}(\alpha,\mu/p).
\end{equation}


\section{Calculation of the matter superfields contribution.}
\hspace{\parindent} \label{Section_New_Identity}

To calculate a contribution of matter superfields it is convenient
to use an approach, based on substituting solutions of
Slavnov--Taylor identities into the Schwinger--Dyson equations.
The Schwinger--Dyson equations in the considered theory can be
written as

\begin{eqnarray}\label{SD1}
&& \frac{\delta\Gamma}{\delta {\bf V}_x} = \frac{1}{2} \Big\langle
\phi_x^* e^{2V_x} \phi_x -\tilde\phi_x^* e^{2V_x} \tilde\phi_x +
\Big(\phi_{0x}^* e^{2V_x} \phi_x -\tilde\phi_{0x}^* e^{2V_x}
\tilde\phi_x +\mbox{h.c.}\Big)
- (PV) \Big\rangle;\qquad\\
\label{SD2} && \frac{\delta\Gamma}{\delta\phi_x^*} =
-\frac{D_x^2}{2}\Big\langle e^{2V_x}\phi_x + e^{2V_x} \phi_{0x}
\Big\rangle = - 2 D_x^2 \frac{\delta\Gamma}{\delta\phi_{0x}^*}
-\frac{D_x^2}{2}\Big\langle e^{2V_x} \phi_{0x} \Big\rangle,
\end{eqnarray}

\noindent where $(PV)$ in the first equation denotes contributions
of the Pauli--Villars fields. Due to the second equation,
derivatives with respect to the additional sources $\phi_0$ and
$\tilde\phi_0$ are very similar to derivatives with respect to the
fields $\phi$ and $\tilde\phi$. The difference of Feynman diagrams
is that there is the operator $\bar D^2$ on external $\phi$-lines,
and there is no this operator on external $\phi_0$-lines.

Let us differentiate the first Schwinger--Dyson equation with
respect to ${\bf V}_y$ and set all fields and sources to 0:

\begin{eqnarray}
\frac{\delta^2\Gamma}{\delta {\bf V}_y \delta {\bf V}_x} =
\frac{2}{i}\frac{\delta}{\delta {\bf V}_y}
\Bigg(\frac{\delta}{\delta j_x^*} \frac{\delta W}{\delta
\phi_{0x}^*} - \frac{\delta}{\delta \tilde j_x^*} \frac{\delta
W}{\delta \tilde\phi_{0x}^*}\Bigg).
\end{eqnarray}

\noindent Then we commute the variational derivative with respect
to the field ${\bf V}_y$ and derivatives with respect to the
sources $j^*$ and $\tilde j^*$. Moreover, we take into account
that the fields $\phi_0^*$ are some parameters. Hence,

\begin{equation}\label{Phi0_Derivative}
\frac{\delta W}{\delta\phi_0^*} =
\frac{\delta\Gamma}{\delta\phi_0^*}.
\end{equation}

\noindent As a result we obtain

\begin{equation}\label{With_Commutators}
\frac{\delta^2\Gamma}{\delta {\bf V}_y \delta {\bf V}_x} =
\frac{2}{i}\Bigg(\Big[\frac{\delta}{\delta {\bf V}_y},
\frac{\delta}{\delta j_x^*}\Big] \frac{\delta\Gamma}{\delta
\phi_{0x}^*} - \Big[\frac{\delta}{\delta {\bf V}_y},
\frac{\delta}{\delta \tilde j_x^*}\Big] \frac{\delta\Gamma}{\delta
\tilde\phi_{0x}^*} + \frac{\delta}{\delta j_x^*} \frac{\delta^2
\Gamma}{\delta \phi_{0x}^* \delta {\bf V}_y} -
\frac{\delta}{\delta \tilde j_x^*} \frac{\delta^2 \Gamma}{\delta
\tilde \phi_{0x}^*\delta {\bf V}_y} \Bigg).
\end{equation}

\noindent Taking into account that the source $j^*$ is an
antichiral field, a contribution of the last two terms to the
effective action can be written as

\begin{eqnarray}\label{Second_Term}
&& -i\int d^8x\,d^8y\,{\bf V}_y {\bf V}_x \frac{\delta}{\delta
j^*_x} \frac{\delta^2\Gamma}{\delta\phi_{0x}^* \delta {\bf V}_y} =
i \int d^8x\,d^8y\,{\bf V}_y {\bf V}_x \frac{D_x^2 \bar
D_x^2}{16\partial^2} \frac{\delta}{\delta j^*_x}
\frac{\delta^2\Gamma}{\delta\phi_{0z}^* \delta {\bf
V}_y}\Bigg|_{z=x} =
\nonumber\\
&& = i \int d^8x\,d^8y\,{\bf V}_y \Bigg({\bf V}_x \frac{\bar
D_x^2}{16\partial^2} \frac{\delta}{\delta j^*_x} D_x^2
\frac{\delta^2\Gamma}{\delta\phi_{0x}^* \delta {\bf V}_y} + D_x^2
{\bf V}_x \frac{\bar D_x^2}{16\partial^2} \frac{\delta}{\delta
j^*_x} \frac{\delta^2\Gamma}{\delta\phi_{0x}^* \delta {\bf V}_y}
+\nonumber\\
&& + D_x^b {\bf V}_x \frac{\bar D_x^2}{8\partial^2}
\frac{\delta}{\delta j^*_x} D_{bx}
\frac{\delta^2\Gamma}{\delta\phi_{0x}^* \delta {\bf V}_y}\Bigg).
\end{eqnarray}

\noindent The first term in this expression and a similar term
with a commutator in Eq. (\ref{With_Commutators}) can be expressed
in terms of the usual Green functions (which do not contain the
additional sources) using the Schwinger--Dyson equations for
matter superfields. They was calculated, for example, in Ref.
\cite{SDYM}. The result is (taking into account a similar
contribution of the fields $\tilde\phi$)

\begin{eqnarray}
&& \frac{d}{d\ln\Lambda}\frac{\delta^2\Gamma}{\delta {\bf V}_y
\delta {\bf V}_x}\Bigg|_{p=0} = \partial^2\Pi_{1/2}
\delta^8_{xy}\, \int\frac{d^4q}{(2\pi)^4}\frac{d}{d\ln\Lambda}
\Bigg\{ \frac{1}{q^2}\frac{d}{dq^2} \ln (q^2 G^2)
-(PV)\Bigg\}+\nonumber\\
&& +\mbox{other contributions},
\end{eqnarray}

\noindent According to Refs. \cite{SD,SDYM}, if all other
contributions to the considered Green function are 0, then the
presented expression corresponds to the exact NSVZ
$\beta$-function.

However, the last two terms in Eq. (\ref{Second_Term}) can not be
already written in terms of the usual Green functions using the
Schwinger--Dyson equations for matter superfields. Moreover, it is
possible to find out \cite{SD,SDYM} that they are expressed in
terms of functions that can not be found from the Slavnov--Taylor
identities. However, explicit calculations show that they are
always 0. Thus, we should propose the existence of a new identity

\begin{eqnarray}
&& \frac{d}{d\ln\Lambda} \int d^8x\,d^8y\,{\bf V}_y \Big(D_x^2
{\bf V}_x \frac{\bar D_x^2}{2\partial^2} \frac{\delta}{\delta
j^*_x} \frac{\delta^2\Gamma}{\delta\phi_{0x}^* \delta {\bf V}_y} +
D_x^b {\bf V}_x \frac{\bar D_x^2}{\partial^2} \frac{\delta}{\delta
j^*_x} D_{bx} \frac{\delta^2\Gamma}{\delta\phi_{0x}^*
\delta {\bf V}_y}\Big)+\nonumber\\
&& +\mbox{similar terms with $\tilde j$ and $\tilde\phi_0$} =0.
\end{eqnarray}

In order to  rewrite these terms in a more visual form, we again
use the Schwinger--Dyson equation  for the gauge field. The result
is

\begin{eqnarray}
&& \frac{d}{d\ln\Lambda} \int d^8x\,d^8y\,{\bf V}_y \Big\langle
\Big( D_x^2 {\bf V}_x \frac{\bar D_x^2}{2\partial^2}\phi^*_x
e^{2V_x} \phi_x + D_x^b {\bf V}_x \frac{D_{xb} \bar
D_x^2}{\partial^2}\phi^*_x e^{2V_x} \phi_x
\Big) \phi^*_y e^{2V_y} \phi_y \Big\rangle +\nonumber\\
&& + \mbox{similar terms with $\tilde\phi$} =0.
\end{eqnarray}

\noindent This equation is a functional form of the new identity
for Green functions. We remark that products of two-point
correlators, which appear deriving this identity, are 0. For
example,

\begin{eqnarray}
&& \frac{1}{4}\Big\langle D^b{\bf V}_x \frac{D_b \bar
D^2}{\partial^2}\phi^*_x e^{2V_x} \phi_x + D^2{\bf
V}_x\frac{\bar D^2}{2\partial^2}\phi^*_x  e^{2V_x} \phi_x\Big\rangle =\nonumber\\
&&\qquad\qquad\qquad\qquad = -i D^b{\bf V}_x \frac{D_{xb}\bar
D_x^2}{\partial^2} \frac{\delta}{\delta j^*_x}
\frac{\delta\Gamma}{\delta\phi_{0z}^*}\Bigg|_{z=x} - i D^2{\bf
V}_x \frac{\bar D_x^2}{2\partial^2} \frac{\delta}{\delta j^*_x}
\frac{\delta\Gamma}{\delta\phi_{0z}^*}\Bigg|_{z=x} =0.\qquad
\end{eqnarray}

\noindent (To verify the last equality, it is possible to express
the derivative with respect to the source in terms of derivatives
with respect to fields and to use explicit expression for the
two-point Green functions. Equality to 0 is obtained, because the
expression $\hat P_x \delta^8_{xy}|_{x=y}$ is not 0 only if the
operator $\hat P$ contains 4 spinor derivatives.) Due to the
similar reasons, the terms obtained, if we differentiate the
fields $\phi_0$ in the Schwinger--Dyson equation for the gauge
field, and the terms containing both $\phi$ and $\tilde\phi$ are
0.

Then, we use the identity

\begin{eqnarray}\label{Identity_With_Derivatives}
&& \phi^* {\bf V} = - \frac{D^2 \bar D^2}{16
\partial^2} \phi^* {\bf V}  = - \frac{D^2 \bar D^2}{16 \partial^2}
(\phi^* {\bf V}) + \frac{1}{16\partial^2} \phi^* D^2 \bar D^2 {\bf
V} + \frac{\bar D^a}{8\partial^2} \phi^* D^2 \bar D_a {\bf V}
+\nonumber\\
&& + \frac{\bar D^2}{16\partial^2} \phi^* D^2 {\bf V} - i
(\gamma^\mu)_{ab} \frac{\partial_\mu}{2\partial^2} \phi^* D^a \bar
D^b {\bf V} + \frac{D^a \bar D^2}{8\partial^2} \phi^* D_a {\bf V}.
\end{eqnarray}

\noindent In our notation $(\gamma^\mu)_{ab}\equiv
-(\gamma^\mu)_a{}^c C_{cb}$, where $(\gamma^\mu)_a{}^c$ are usual
$\gamma$-matrices, and $C$ is a charge conjugation matrix.

In Appendix \ref{Appendix_0} we show that the first term in this
expression does not contribute to the new identity. Some other
terms are also 0. To verify this, we note that any third power of
the chiral derivative can be written as an expression containing
the derivative $\partial_\mu$, which does not act on the
background field ${\bf V}$, because we consider the limit of zero
external momentum. Therefore, the new identity can be rewritten as

\begin{eqnarray}\label{New_Identity}
&& \frac{d}{d\ln\Lambda} \int d^8x\,d^8y\,\Big(D^a {\bf V}_x i
(\gamma^\mu)_{bc} D^b \bar D^c {\bf V}_y \Big\langle \frac{D_a
\bar D^2}{\partial^2}\phi^*_x e^{2V_x} \phi_x
\frac{\partial_\mu}{2\partial^2} \phi_y^* e^{2V_y} \phi_y
\Big\rangle +\qquad \nonumber\\
&& + D^a {\bf V}_x D^{b} {\bf V}_y \Big\langle \frac{D_a \bar
D^2}{\partial^2}\phi^*_x e^{2V_x} \phi_x \frac{D_b \bar
D^2}{8\partial^2} \phi_y^* e^{2V_y} \phi_y \Big\rangle\Big) = 0.
\end{eqnarray}

\noindent The next section of the paper is devoted to a possible
way for proving this identity.


\section{Way of proving the new identity}
\hspace{\parindent}\label{Section_Proof}

Now, let us describe how one can try to prove the new identity for
Green functions. Actually the idea was formulated in Ref.
\cite{Identity}, analyzing the Feynman rules. Here we will use
strict functional methods. Moreover, there is a mistake in a sign
in Ref. \cite{Identity}. To correct this error, it is necessary to
slightly modify the proof.

It is convenient to write the new identity in form
(\ref{New_Identity}). The functional integral over the matter
fields is Gaussian and can be calculated explicitly. We will need
the equation

\begin{eqnarray}\label{Weak_Theorem}
&& \int dx_1\ldots dx_n\,x_a x_b x_c x_d \exp\Big(-\frac{1}{2} x_i
A_{ij} x_j\Big) =\nonumber\\
&&\qquad\qquad\qquad\qquad = \mbox{const}\,(\det A)^{-1/2}
\Big(A_{ab}^{-1} A_{cd}^{-1} + A_{ac}^{-1} A_{bd}^{-1} +
A_{ad}^{-1} A_{bc}^{-1} \Big).\qquad
\end{eqnarray}

\noindent In the massless case $A$ is the operator

\begin{equation}
D^2 e^{2V} \bar D^2 = D^2 \bar D^2 + D^2 (e^{2V}-1) \bar D^2.
\end{equation}

\noindent The operator inverse to $A$ by definition satisfies the
condition

\begin{equation}
\frac{1}{D^2 e^{2V} \bar D^2} D^2 e^{2V} \bar D^2 =D^2 e^{2V} \bar
D^2 \frac{1}{D^2 e^{2V} \bar D^2} = - \frac{D^2 \bar
D^2}{16\partial^2}\equiv \hat 1.
\end{equation}

\noindent It can be easily constructed explicitly:

\begin{eqnarray}\label{Inverse_Operator}
&& \frac{1}{D^2 e^{2V} \bar D^2} = \Big(1 - \frac{1}{16\partial^2}
D^2 (e^{2V}-1) \bar
D^2\Big)^{-1}\frac{D^2\bar D^2}{16^2\partial^4} =\nonumber\\
&&\qquad\qquad\qquad\qquad\qquad\qquad = \sum\limits_{n=0}^\infty
\Big(\frac{1}{16\partial^2} D^2 (e^{2V}-1) \bar D^2\Big)^n
\frac{D^2\bar D^2}{16^2\partial^4}.\qquad
\end{eqnarray}

\noindent After the functional integration over matter superfields
by using Eq. (\ref{Weak_Theorem}), the left hand side of the new
identity (up to an insignificant factor) becomes

\begin{eqnarray}\label{Identity_LHS}
&&  \mbox{Tr}\, \frac{d}{d\ln\Lambda} \Big\langle \frac{1}{D^2
e^{2V} \bar D^2} D^a (D_a {\bf V}) e^{2V} \bar D^2
\Big(i(\gamma_\mu)_{bc} \frac{1}{D^2 e^{2V} \bar D^2}
\frac{\partial_\mu}{\partial^2} D^2 (\bar D^b D^c {\bf V})
e^{2V}\bar D^2 -\nonumber\\
&& - \frac{4}{D^2 e^{2V}\bar D^2} D^b (D_b {\bf V}) e^{2V}\bar
D^2\Big)\Big\rangle.
\end{eqnarray}

\noindent The trace includes the integration over the superspace:

\begin{equation}
\mbox{Tr} \hat A \equiv \int d^8x\,A_{xx},
\end{equation}

\noindent and the angular brackets here and below denote taking
the vacuum expectation value by the functional integration only
with respect to the gauge field.

We remark that it is also possible to use a brief notation
introduced in Ref. \cite{Identity}:

\begin{eqnarray}\label{Notation}
&& * \equiv -\frac{4}{D^2 e^{2V}\bar D^2};\qquad\ \  (\bar I_1)^a
\equiv \frac{1}{2} D^2 e^{2V} \bar D^a; \qquad\ \, (I_1)^a \equiv
\frac{1}{2} D^a e^{2V} \bar D^2;
\nonumber\\
&& (I_2) \equiv \frac{1}{4} e^{2V} \bar D^2;\qquad\quad\, (\bar
I_2) \equiv \frac{1}{4} D^2 e^{2V};\qquad\qquad\  (I_2)^{ab}
\equiv D^a e^{2V} \bar D^b;\nonumber\\
&& (I_3)^a \equiv \frac{1}{2} D^a e^{2V}; \qquad\ \ \, (\bar
I_3)^a \equiv \frac{1}{2} e^{2V} \bar D^a;\qquad\quad\ \ \, (I_0)
\equiv \frac{1}{4} D^2 e^{2V} \bar D^2.
\end{eqnarray}

\begin{figure}[h]
\hspace*{2.9cm}
\begin{picture}(0,0)
\put(-0.2,0.1){$D_a {\bf V}$} \put(3.2,0.1){$D^b {\bf V}$}
\put(2.4,-0.3){$(I_1)_b$} \put(0.7,1.3){$(I_1)^a$}
\put(5.2,0.1){$D_a {\bf V}$} \put(8.6,0.9){$\bar D
\gamma^\mu\gamma_5 D {\bf V}$} \put(7.8,-0.3){$q_\mu/4q^2 (I_0)$}
\put(6.1,1.3){$(I_1)^a$} \put(4.4,0.6){$+$}
\end{picture}
\includegraphics[scale=0.4]{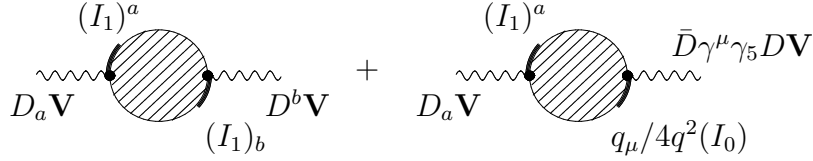}
\hspace*{2cm}
\includegraphics[scale=0.4]{ins3.eps}
\newline
\caption{Graphical form of Eq. (\ref{Identity_LHS}).}
\label{Figure_X}
\end{figure}

\noindent In Ref. \cite{Identity} these expressions were defined
differently, since in that paper we did not use functional
methods. Nevertheless, main formulas, strictly derived in this
paper, are similar to formulas in \cite{Identity}. Using this
notation, Eq. (\ref{Identity_LHS}) can be graphically presented as
a sum of two effective diagrams, presented in Fig. \ref{Figure_X}.

If the new identity is true, than expression (\ref{Identity_LHS})
is 0. In order to verify this, we have to commute factors
containing the background field with the expression $(D^2
e^{2V}\bar D^2)^{-1}$. This can be made by using the identities

\begin{eqnarray}
&& \hat 1\cdot [A^{-1},X] = A^{-1} [X,A] A^{-1} + A^{-1}
[\hat 1,X];\nonumber\\
&& [A^{-1},X]\cdot \hat 1 = A^{-1} [X,A] A^{-1} + [\hat 1,X]
A^{-1}.
\end{eqnarray}

\noindent If $X_{yz} = X_y \delta^8_{yz}$ and $A^{-1}_{xy} = \hat
O_x \delta^8_{xy}$, where $\hat O$ is an operator, then

\begin{equation}
[A^{-1},X]_{xz} = A^{-1}_{xz} X_z - X_x A^{-1}_{xz} = \hat O_x
\delta^8_{xz} X_z - X_x \hat O_x \delta^8_{xz} = \hat O_x(X_x
\delta^8_{xz}) -X_x \hat O_x \delta^8_{xz}.
\end{equation}

\noindent To calculate commutators with other expressions
containing the supersymmetric covariant derivatives, below we will
use the following identities:

\begin{eqnarray}\label{Commutators}
&& \bar D^2 X = X \bar D^2 - 2 (-1)^{P_x} (\bar D_a X) \bar D^a +
(\bar D^2 X);\nonumber\\
&& X D^2 = D^2 X - 2 D^a (D_a X) + (D^2 X);\nonumber\\
&& \bar D^a X = (-1)^{P_X} X \bar D^a + (\bar D^a X);\nonumber\\
&& X D^a = (-1)^{P_X} D^a X - (-1)^{P_X} (D^a X),
\end{eqnarray}

\noindent where $P_X$ denotes the Grassmanian parity of the
expression $X$, which is 0(mod 2) or 1(mod 2).

Now let us proceed to calculating expression (\ref{Identity_LHS}).
For this purpose we note that acting on ${\bf V}$ the derivatives
$\partial_\mu$ gives 0, because the momentum of this field is 0.
Due to the same reason, more than 4 spinor derivatives can not act
on the field ${\bf V}$. Moreover, the new identity contributes
only to the transversal part of the gauge field Green function.
(This follows from the Ward identity and the result of calculating
two-point Green function using the Schwinger--Dyson equations.)
Hence, we may keep only terms with 2 derivatives $D$ and 2
derivatives $\bar D$ acting on the background field.

Therefore, to simplify the calculations, it is possible to
substitute formally the background field ${\bf V}$ for
$\bar\theta^2 \theta^2$. In this case the calculations are
automatically made in the limit of zero external momentum, because
this expression is independent of the coordinates. Since

\begin{equation}
\int d^4\theta\,(\bar\theta^2 \theta^2) \partial^2\Pi_{1/2}
(\bar\theta^2 \theta^2) = -\frac{1}{2}\int d^4\theta\,\bar\theta^2
\theta^2 = -2,
\end{equation}

\noindent after this substitution

\begin{equation}
\int \frac{d^4p}{(2\pi)^4}\,d^4\theta\,{\bf V}\partial^2\Pi_{1/2}
{\bf V} \int \frac{d^4q}{(2\pi)^4} \frac{f}{q^2 G} \to -2\int
\frac{d^4p}{(2\pi)^4}\,\frac{d^4q}{(2\pi)^4} \frac{f}{q^2 G}.
\end{equation}

\noindent Thus, having obtained formally the right hand side, it
is possible to uniquely restore the left hand side. These
speculations can be easily verified by calculating the simplest
Feynman diagrams, for example, one- or two-loop diagrams.

Using the described method, it is easy to prove that the first
term in Eq. (\ref{Identity_LHS}) is 0. Really, after the
substitution ${\bf V} \to \theta^2 \bar\theta^2$ it is written as

\begin{equation}\label{T1}
T_1 \equiv i(\gamma_\mu)_{bc}  \mbox{Tr}\, \frac{d}{d\ln\Lambda}
\Big\langle \frac{1}{D^2 e^{2V} \bar D^2} D^a \theta_a
\bar\theta^2 e^{2V} \bar D^2 \frac{1}{D^2 e^{2V} \bar D^2}
\frac{\partial_\mu}{\partial^2} D^2 \bar\theta^b \theta^c
e^{2V}\bar D^2\Big\rangle
\end{equation}

\noindent up to an insignificant constant. The derivative with
respect to $\ln\Lambda$ in this equation acts only on propagators
of the gauge superfields, which are obtained calculating the
functional integrals over the fields $V$. Due to the presence of
this derivative, all considered expression are well defined.
Really, in the momentum representation we first make the
differentiation with respect to $\ln\Lambda$, and then perform
integration with respect to loop momentums. Since trace and
differentiation with respect to $\ln\Lambda$ commute, the operator
$\mbox{Tr}\, d/d\ln\Lambda$ is invariant under cyclic
replacements.

We note that the right spinor $\theta$ in Eq. (\ref{T1}) can be
moved in any place. Really, calculating the trace we obtain

\begin{equation}\label{P_Delta}
\int d^4x\,d^4\theta\,\hat P
\delta^8_{12}\Big|_{\theta_1=\theta_2=\theta;\,x_1=x_2=x},
\end{equation}

\noindent where the operator $\hat P$ is a product of covariant
derivatives and also two factors $\theta_a \bar\theta^2$ and
$\bar\theta^b \theta^c$, which are contained in the considered
trace. Commuting $\theta$ and $\bar\theta$ with covariant
derivatives, it is possible to shift all $\theta$ and $\bar\theta$
so that they will be on the left of the covariant derivatives. It
is well known that the action of the spinor covariant derivatives
on $\delta^4(\theta_1-\theta_2)$ in coinciding points is an
expression independent of $\theta$. (In order to verify this, it
is necessary to make a number of the spinor derivatives less than
or equal to 4 using commutation relations. If a number of the
derivatives is less than 4, then acting on the $\delta$-function
in coinciding points gives 0, and if this number is 4, this gives
an expression independent of $\theta$.) Therefore, a nontrivial
contribution to Eq. (\ref{P_Delta}) can be obtained only if there
is the expression

\begin{equation}
\int d^4\theta\,\theta^2 \bar\theta^2 = 4
\end{equation}

\noindent on the left of the covariant derivatives. But the
anticommutator of the covariant derivative with $\theta$ is a
constant. Therefore, all terms with commutators are 0, because a
degree of $\theta$ in them is less than 2. This means that the
right spinor $\theta$ in the considered expression can be moved in
any place. In particular, both $\theta$ can be put together. Then,
using the identity $\theta^c \theta_a = \delta^c_a\,\theta^2/2$
and taking into account that $D$ and $\bar\theta^b$ anticommute,
the considered expression can be rewritten as

\begin{eqnarray}
&& T_1 = \frac{i}{2}(\gamma_\mu)_{ab}
\mbox{Tr}\,\frac{d}{d\ln\Lambda} \Big\langle \frac{1}{D^2 e^{2V}
\bar D^2} D^a \theta^2 \bar\theta^2 e^{2V} \bar D^2 \frac{1}{D^2
e^{2V} \bar D^2} \frac{\partial_\mu}{\partial^2} \bar\theta^b D^2
e^{2V}\bar
D^2\Big\rangle =\nonumber\\
&& = -\frac{i}{2}(\gamma_\mu)_{ab}
\mbox{Tr}\,\frac{d}{d\ln\Lambda} \Big\langle \frac{D^2 \bar
D^2}{16\partial^2} D^a \theta^2 \bar\theta^2 e^{2V} \bar D^2
\frac{1}{D^2 e^{2V} \bar D^2} \frac{\partial_\mu}{\partial^2}
\bar\theta^b \Big\rangle.
\end{eqnarray}

\noindent Moving $\bar\theta^b$ using commutation relations
(\ref{Commutators}) to the right until it will be multiplied by
$\bar\theta^2$, we find

\begin{eqnarray}
&& T_1 = - i(\gamma_\mu)_{ab}  \mbox{Tr}\,\frac{d}{d\ln\Lambda}
\Big\langle \frac{D^2 \bar D^b}{16\partial^2} D^a \theta^2
\bar\theta^2 e^{2V} \bar D^2 \frac{1}{D^2 e^{2V} \bar D^2}
\frac{\partial_\mu}{\partial^2} \Big\rangle = \nonumber\\
&&\qquad\qquad\qquad\qquad\qquad = -
\mbox{Tr}\,\frac{d}{d\ln\Lambda} \Big\langle \frac{\partial^\mu}{8
\partial^2} D^2 \theta^2 \bar\theta^2 e^{2V} \bar D^2 \frac{1}{D^2
e^{2V} \bar D^2} \frac{\partial_\mu}{\partial^2}
\Big\rangle.\qquad
\end{eqnarray}

\noindent We note that $\theta^2 \bar\theta^2$ can be shifted to
an arbitrary place of this expression, since a commutation of
$\theta$ or $\bar\theta$ with the supersymmetric covariant
derivatives decreases a degree of $\theta$ or $\bar\theta$ on 1,
and the integral over $d^4\theta$ is not 0 only if it acts on the
fourth degree. Hence, the first term in Eq. (\ref{Identity_LHS})
can be rewritten as

\begin{eqnarray}
&& T_1 = -  \mbox{Tr}\, \frac{d}{d\ln\Lambda} \Big\langle \theta^2
\bar\theta^2 \frac{\partial^\mu}{8\partial^2} D^2 e^{2V} \bar D^2
\frac{1}{D^2 e^{2V} \bar D^2} \frac{\partial_\mu}{\partial^2}
\Big\rangle =\nonumber\\
&&\qquad\qquad\qquad\qquad\qquad\qquad\qquad = \mbox{Tr}\,
\frac{d}{d\ln\Lambda} \Big\langle\theta^2 \bar\theta^2
\frac{\partial^\mu}{16
\partial^2} \frac{D^2\bar D^2}{8\partial^2}
\frac{\partial_\mu}{\partial^2} \Big\rangle = 0,\qquad
\end{eqnarray}

\noindent where we take into account that the last expression is
independent of $\Lambda$ and disappears after differentiating with
respect to $\ln\Lambda$. Really, in the momentum representation
this expression is proportional to

\begin{equation}
\int \frac{d^4q}{(2\pi)^4} \frac{d}{d\ln\Lambda} \frac{1}{q^4} =
0.
\end{equation}

\noindent Presence of the derivative with respect to $\ln\Lambda$
guarantees that this expression is well defined. (We take the
trace after the differentiation.)

Similarly we can try to prove that the second term in Eq.
(\ref{Identity_LHS}) is 0. However, the corresponding calculation
is much more complicated. First, as earlier we make the formal
substitution ${\bf V} \to \theta^2 \bar\theta^2$, after which the
second term in Eq. (\ref{Identity_LHS}) will be proportional to

\begin{eqnarray}
T_2 \equiv \mbox{Tr}\, \frac{d}{d\ln\Lambda}\Big\langle
\frac{1}{D^2 e^{2V} \bar D^2} D^a \theta_a \bar\theta^2 e^{2V}\bar
D^2 \frac{1}{D^2 e^{2V} \bar D^2} D^b \theta_b \bar\theta^2 e^{2V}
\bar D^2\Big\rangle.
\end{eqnarray}

\noindent Similar to calculating the first term, we can put
together the right spinors $\theta$:

\begin{eqnarray}
T_2 = -\frac{1}{2} \mbox{Tr}\, \frac{d}{d\ln\Lambda}\Big\langle
\frac{1}{D^2 e^{2V} \bar D^2} D^a \theta^2 \bar\theta^2 e^{2V}
\bar D^2 \frac{1}{D^2 e^{2V} \bar D^2} D_a \bar\theta^2 e^{2V}
\bar D^2\Big\rangle.
\end{eqnarray}

\noindent We also put together the factors $\bar\theta$, commuting
them with the covariant derivatives. After some simple
transformations the result can be written as

\begin{eqnarray}\label{T2}
&& T_2 = \mbox{Tr}\,\frac{d}{d\ln\Lambda}\,\theta^2 \bar\theta^2
\Big\langle (\gamma^\mu)_{ab}\frac{i\partial_\mu}{2\partial^2} D^a
e^{2V} \bar D^b \frac{1}{D^2 e^{2V} \bar D^2} - (\gamma^\mu)_{ab}
\frac{i\partial_\mu}{2\partial^2} D^a e^{2V} D^2 \frac{1}{D^2
e^{2V} \bar D^2} \times\nonumber\\
&& \times D^2 e^{2V} \bar D^b \frac{1}{D^2 e^{2V} \bar D^2}
+\frac{1}{D^2 e^{2V} \bar D^2} D^a e^{2V} \frac{1}{D^2 e^{2V}\bar
D^2} D_a e^{2V} \bar D^2
-\nonumber\\
&& - \frac{1}{D^2 e^{2V} \bar D^2} D^a e^{2V} \bar D^2
\frac{1}{D^2 e^{2V}\bar D^2} D^2 e^{2V} \frac{1}{D^2 e^{2V}\bar
D^2} D_a e^{2V} \bar D^2
+\nonumber\\
&& +2 \frac{1}{D^2 e^{2V} \bar D^2} D^a e^{2V} \bar D^b
\frac{1}{D^2 e^{2V}\bar D^2} D^2 e^{2V}\bar D_b \frac{1}{D^2
e^{2V}\bar D^2} D_a e^{2V} \bar D^2
-\\
&& -2 \frac{1}{D^2 e^{2V} \bar D^2} D^a e^{2V} \bar D^2
\frac{1}{D^2 e^{2V}\bar D^2}D^2 e^{2V}\bar D^b \frac{1}{D^2
e^{2V}\bar D^2} D^2 e^{2V}\bar D_b\frac{1}{D^2 e^{2V}\bar D^2} D^b
e^{2V} \bar D^2 \Big\rangle.\nonumber
\end{eqnarray}

\noindent Here we take into account that all terms, in which an
overall degree of $\theta$ and $\bar\theta$ is not 4, disappear
after integrating over the anticommuting variables. In particular,
due to this reason, the expression $\theta^2 \bar\theta^2$ in the
last identity can be moved to an arbitrary place of the trace. The
sum of 2 first terms in Eq. (\ref{T2}) is 0, because it is
proportional to

\begin{eqnarray}
&& \mbox{Tr}\,\frac{d}{d\ln\Lambda}\,\theta^2 \bar\theta^2
\Big[\bar\theta^b,\, \Big\langle
(\gamma^\mu)_{ab}\frac{i\partial_\mu}{4\partial^2} D^a e^{2V} \bar
D^2 \frac{1}{D^2 e^{2V} \bar D^2}\Big\rangle\Big] +
\mbox{Tr}\,\frac{d}{d\ln\Lambda}\,\theta^2 \bar\theta^2
\Big\langle (\gamma^\mu)_{ab}\frac{i\partial_\mu}{4\partial^2}
\times\nonumber\\
&& \times D^a e^{2V} \bar D^2 \frac{1}{D^2 e^{2V} \bar D^2}
\frac{D^2 \bar D^b}{16\partial^2} \Big\rangle = -
\mbox{Tr}\,\frac{d}{d\ln\Lambda}\,\theta^2 \bar\theta^2 \frac{D^2
\bar D^2}{16^2\partial^2} = 0.
\end{eqnarray}

\noindent Using this equality, in brief notation (\ref{Notation})
the expression for $T_2$ can be rewritten in the more compact
form:

\begin{eqnarray}\label{Considered_Terms}
&& T_2 = \frac{1}{8} \frac{d}{d\ln\Lambda}
\mbox{Tr}\,\theta^2\bar\theta^2 \Big\langle 2 * (I_3)^a * (I_1)_a
+ 2  * (\bar I_2) * (I_1)^a * (I_1)_a - * (I_2)^{ab} * (\bar
I_1)_b * (I_1)_a
-\vphantom{\frac{1}{2}}\nonumber\\
&& - * (I_1)^a * (\bar I_1)^b * (\bar I_1)_{b} * (I_1)_a
\Big\rangle.
\end{eqnarray}

To simplify it, we consider the trace of the commutator

\begin{eqnarray}
&& 0 = (\gamma^\mu)_{ab}
\mbox{Tr}\frac{d}{d\ln\Lambda}\,\Big[y^\mu, \theta^2 \bar\theta^2
\Big\langle \frac{1}{D^2 e^{2V}\bar D^2} D^a
e^{2V} \bar D^b \Big\rangle \Big] =\nonumber\\
&&\qquad\qquad = (\gamma^\mu)_{ab}
\mbox{Tr}\frac{d}{d\ln\Lambda}\,\Big[y^\mu, \theta^2 \bar\theta^2
\Big\langle \frac{D^2\bar D^2}{16\partial^2} \frac{1}{D^2
e^{2V}\bar D^2} \frac{D^2\bar D^2}{16\partial^2} D^a e^{2V} \bar
D^b \Big\rangle \Big],\qquad
\end{eqnarray}

\noindent where $y^\mu \equiv x^\mu +
i\bar\theta\gamma^\mu\gamma_5 \theta/2$. (In the momentum
representation this expression is evidently an integral of a total
derivative with respect to the loop momentum.) For calculating
this expression we will first use the identities

\begin{equation}
[y^\mu, \bar D_a] = 0;\qquad [y^\mu, D_a] = 2i(\gamma^\mu
\bar\theta)_a.
\end{equation}

\noindent We obtain

\begin{eqnarray}
&& 0 = (\gamma^\mu)_{ab} \mbox{Tr}\frac{d}{d\ln\Lambda}\, \theta^2
\bar\theta^2 \Big\langle \frac{2i}{D^2 e^{2V}\bar D^2}
(\gamma^\mu\bar\theta)^a e^{2V} \bar D^b - \frac{4i}{D^2
e^{2V}\bar D^2} (\gamma^\mu\bar\theta)^c D_c e^{2V} \bar D^2
\frac{1}{D^2 e^{2V}\bar D^2}
\times\qquad\nonumber\\
&& \times D^a e^{2V} \bar D^b + (\gamma^\mu)_{ab}
\frac{2\partial_\mu}{\partial^2}\frac{1}{D^2 e^{2V}\bar D^2} D^a
e^{2V} \bar D^b \Big\rangle = \mbox{Tr}\frac{d}{d\ln\Lambda}\,
\theta^2 \bar\theta^2 \Big\langle -\frac{8i}{D^2 e^{2V}\bar D^2}
\bar\theta_b e^{2V} \bar D^b +\nonumber\\
&& + \frac{8i}{D^2 e^{2V}\bar D^2} \bar\theta_b D_a e^{2V}\bar D^2
\frac{1}{D^2 e^{2V}\bar D^2} D^a e^{2V} \bar D^b +
(\gamma^\mu)_{ab}\frac{2\partial_\mu}{\partial^2}\frac{1}{D^2
e^{2V}\bar D^2} D^a e^{2V} \bar D^b
\Big\rangle.\vphantom{\frac{1}{2}}
\end{eqnarray}

\noindent Then, we put together all $\theta$ and $\bar\theta$,
commuting them with the covariant derivatives:

\begin{eqnarray}
&& 0 = \mbox{Tr}\frac{d}{d\ln\Lambda}\, \theta^2 \bar\theta^2
\Big\langle \frac{1}{D^2 e^{2V}\bar D^2} D^2 e^{2V}\bar D^a
\frac{1}{D^2 e^{2V}\bar D^2} e^{2V} \bar D_a +\nonumber\\
&&\qquad\qquad + \frac{1}{D^2 e^{2V}\bar D^2} D^2 e^{2V}\bar D_b
\frac{1}{D^2 e^{2V}\bar D^2} D_a e^{2V}\bar D^2 \frac{1}{D^2
e^{2V}\bar D^2} D^a e^{2V} \bar D^b \Big\rangle.\qquad
\end{eqnarray}

\noindent In the brief notation this equality can be rewritten as

\begin{equation}\label{Equality1}
\mbox{Tr}\,\frac{d}{d\ln\Lambda}\,\theta^2 \bar\theta^2
\Big\langle 4 * (\bar I_3)^a * (\bar I_1)_a - * (I_2)^{ab} * (\bar
I_1)_b * (I_1)_a\Big\rangle = 0.
\end{equation}

\noindent Taking into account that, evidently,

\begin{equation}
\mbox{Tr}\,\frac{d}{d\ln\Lambda}\,\theta^2 \bar\theta^2
\Big\langle * (\bar I_3)^a * (\bar I_1)_a \Big\rangle =
\mbox{Tr}\,\frac{d}{d\ln\Lambda}\,\theta^2 \bar\theta^2
\Big\langle * (I_3)^a * (I_1)_a \Big\rangle,
\end{equation}

\noindent the result for the considered correlator can be
simplified:

\begin{eqnarray}\label{Final_T2}
&& T_2 = \frac{1}{8}
\mbox{Tr}\frac{d}{d\ln\Lambda}\,\theta^2\bar\theta^2 \Big\langle 2
* (\bar I_2) * (I_1)^a * (I_1)_a - \frac{1}{2}
* (I_2)^{ab} * (\bar I_1)_b * (I_1)_a -\nonumber\\
&& - * (I_1)^a * (\bar I_1)^b * (\bar I_1)_{b} * (I_1)_a
\Big\rangle.
\end{eqnarray}

\noindent Unfortunately, we could not so far prove that this
expression was 0. This is a key point in the proof of the new
identity and, therefore, of the exact NSVZ $\beta$-function.
Possibly, this expression is not an integral of total derivatives,
but due to some reasons does not contribute to the three- and
four-loop diagrams, which were calculated earlier. Then, it is
necessary to modify the exact $\beta$-function by the following
way:

\begin{equation}
\beta(\alpha) = \frac{\alpha^2}{\pi}
\Big(1-\gamma(\alpha)-\delta(\alpha) \Big),
\end{equation}

\noindent where the function $\delta(\alpha)$ is defined as
follows: If we define the operator

\begin{equation}
\hat O^a \equiv Z \frac{D^a \bar D^2}{2\partial^2}\phi^*
e^{2V}\phi
\end{equation}

\noindent and write its correlator in the form

\begin{equation}
\langle\hat O_x^a\,\hat O^b_{y} \rangle = -\frac{i}{2\pi^2} C^{ab}
\Delta(\alpha,\partial^2/\mu^2) \bar D^2 \delta^8_{xy},
\end{equation}

\noindent then

\begin{equation}
\delta\Big(d(\alpha,\mu/p)\Big) \equiv \frac{\partial}{\partial\ln
p} \Delta(\alpha,\mu/p),
\end{equation}

\noindent where the function $d$ is defined by Eq.
(\ref{D_Definition}).


\section{Conclusion}
\label{Section_Conclusion} \hspace{\parindent}

In this paper we investigate an exact expression for the
contribution of matter superfields to the Gell-Mann--Low function
for the massless $N=1$ supersymmetric electrodynamics. This
investigation is based on substituting solutions of the
Slavnov--Taylor identities into the Schwinger--Dyson equations. In
particular, for contributions of matter superfields we try to
prove an interesting feature, which was first noted in Ref.
\cite{3LoopHEP}: in supersymmetric theories the Gell-Mann--Low
function is given by integrals of total derivatives. This is a
fact that is responsible for the new identity, which was first
proposed in Ref. \cite{SD}. This identity appears, if we require
that the exact Gell-Mann--Low function coincides with the exact
NSVZ $\beta$-function, and does not follow from any known symmetry
of the theory. In this paper we tried to strictly derive the new
identity for the massless case in the Abelian theory. The key
point of the proof is a functional formulation of the new
identity, proposed in Ref. \cite{Review}. Actually, the
calculations, presented here, repeat qualitative speculations of
Ref. \cite{Identity}, but there are essential differences in some
points. Using the proposed method, it is possible to prove that
some contributions to the new identity are really integrals of
total derivatives and equal to 0. However, there are some
contributions, which we could not not factorize to total
derivatives. In principle, there are 2 possibilities: either the
proof should be made differently and the Gell-Mann--Low function
coincides with the exact NSVZ $\beta$-function, or the situation
is similar to

\begin{equation}
\frac{987654321}{123456789} \approx 8.000000073,
\end{equation}

\noindent i.e. the factorization into total derivatives is an
accidental fact, appearing only in the lowest loops, and real
structure of the result becomes clear only in the four-loop or
higher approximation. Then, it is necessary to modify the
expression for the $\beta$-function and add a contribution from
the correlator of some composite operator.

If the new identity is true, then it is important to find a
symmetry, responsible for the new identity. Existence of new
symmetries was earlier proposed investigating finite $N=1$
supersymmetric theories \cite{Ermushev}. Possibly, these
symmetries are somehow related with the AdS/CFT-correspondence
\cite{Maldacena}, but so far they are not yet constructed.
Nevertheless, the formulation of the new identity in terms of
correlators for some composite operators suggests that these
operators corresponds to fields in another theory. Then the
equality of their correlator to 0 can follow from some symmetry of
this theory.

\bigskip
\bigskip

\noindent {\Large\bf Acknowledgments.}

\bigskip

\noindent This paper was supported by the Russian Foundation for
Basic Research (Grant No. 05-01-00541).


\appendix

\section{Appendix: Why the contribution of the first term in Eq.
(\ref{Identity_With_Derivatives}) is 0}.\label{Appendix_0}

It is easy to see that the contribution of the first term in Eq.
(\ref{Identity_With_Derivatives}) to the new identity is 0.
Really, let us consider, for example,

\begin{eqnarray}
&& \frac{d}{d\ln\Lambda} \int d^8x\,d^8y\,\Big\langle D_x^2 {\bf
V}_x \frac{\bar D_x^2}{\partial^2} \phi_x^*
e^{2V_x}\phi_x\,\frac{D_y^2 \bar D_y^2}{\partial^2}(\phi_y^* {\bf
V}_y) e^{2V_y}\phi_y\Big\rangle =\nonumber\\
&& = \frac{d}{d\ln\Lambda} \int d^8x\,d^8y\,D_x^2 {\bf V}_x {\bf
V}_y \Big\langle \frac{\bar D_x^2}{\partial^2} \phi_x^*
e^{2V_x}\phi_x\,\phi_y^* \frac{\bar D_y^2 D_y^2}{\partial^2}
(e^{2V_y}\phi_y)\Big\rangle =\nonumber\\
&& = 16i\frac{d}{d\ln\Lambda} \int d^8x\,d^8y\,D_x^2 {\bf V}_x
{\bf V}_y \frac{\bar D_x^2}{\partial^2}\frac{\delta}{\delta j_x^*}
\frac{\delta}{\delta\phi_{0z}^*} \frac{\delta}{\delta j_y^*}
\frac{\bar D_y^2 D_y^2}{\partial^2} \frac{\delta
W}{\delta\phi_{0y}^*}\Bigg|_{z=x},
\end{eqnarray}

\noindent because, as earlier, all terms with two-point
correlators are 0. (All two-point correlators are expressed in
terms of the only function $G$, and this function is cancelled in
their products. Hence, they disappear after differentiating with
respect to $\ln\Lambda$.)

Now, let us prove that the last four-point correlator is 0. For
this purpose we first use Eq. (\ref{Phi0_Derivative}), and, then,
express the derivative with respect to the additional source
$\phi_0^*$ in terms of the derivative with respect to the field
$\phi$, using Schwinger--Dyson equation for the matter superfield
(\ref{SD2}). We obtain

\begin{eqnarray}
&& \int d^8x\,d^8y\,D_x^2 {\bf V}_x {\bf V}_y
\frac{D_x^2}{\partial^2}\frac{\delta}{\delta j_x^*}
\frac{\delta}{\delta\phi_{0z}^*} \frac{\delta}{\delta j_y^*}
\frac{\bar D_y^2 D_y^2}{\partial^2} \frac{\delta
W}{\delta\phi_{0y}^*}\Bigg|_{z=x} = -\frac{1}{2}\int
d^8x\,d^8y\,D_x^2 {\bf V}_x {\bf V}_y \times\nonumber\\
&& \times \frac{D_x^2}{\partial^2}\frac{\delta}{\delta j_x^*}
\frac{\delta}{\delta\phi_{0z}^*} \frac{\delta}{\delta j_y^*}
\frac{\bar D_y^2}{\partial^2} \frac{\delta \Gamma}{\delta\phi_y^*}
\Bigg|_{z=x} = \int d^8x\,d^8y\,D_x^2 {\bf V}_x {\bf V}_y
\frac{D_x^2}{\partial^2}\frac{\delta}{\delta j_x^*}
\frac{\delta}{\delta\phi_{0z}^*}
\frac{1}{\partial^2}\delta^4_{yw}\Big|_{z=x,w=y} = 0,\qquad
\end{eqnarray}

\noindent because

\begin{equation}
\frac{\delta\Gamma}{\delta\phi^*} = - j^*
\end{equation}

\noindent even if fields and sources are not 0.

Completely similarly we obtain

\begin{equation}
\int d^8x\,d^8y\,\Big\langle D_x^b {\bf V}_x \frac{D_{xb}
D_x^2}{\partial^2} \phi_x^* e^{2V_x}\phi_x\,\frac{D_y^2 \bar
D_y^2}{\partial^2}(\phi_y^* {\bf V}_y) e^{2V_y}\phi_y\Big\rangle =
0.
\end{equation}


\end{document}